\documentclass[aps,prl,twocolumn,groupedaddresss,showpacs]{revtex4}

\newcommand{\ben}{\begin{enumerate}}
\newcommand{\een}{\end{enumerate}}
\newcommand{\be}{\begin{equation}}
\newcommand{\ee}{\end{equation}}
\newcommand{\bea}{\begin{eqnarray}}
\newcommand{\eea}{\end{eqnarray}}

   \def\br{\mbox{\boldmath $r$}}
   \def\bR{\mbox{\boldmath $R$}}

   \def\bnu{\mbox{\boldmath $\nu$}}
   \def\pd{\partial}

\begin{document}

\title{Integral Representations for Free Energies of Macroionic Suspensions
and Equation of State for Osmotic Pressure}

\author{Ikuo S. Sogami}
\email[]{sogami@cc.kyoto-su.ac.jp}
\affiliation{Department of Physics, Kyoto Sangyo University, Kyoto 603-8555, Japan}
\date{\today}

\begin{abstract}
A generating functional which results in the Poisson-Boltzmann equation and boundary conditions for an average electric potential of a macroionic suspension through an extremal condition is constructed in a mean field theory. The extremum of the generating functional turns out to be identical with the Helmholtz free energy of the system which has an integral representation in terms of the average electric potential satisfying the Poisson-Boltzmann equation. From the Helmholtz free energy, the chemical potentials of small ions and {\it chemical potentials of effective valencies of macroions} are calculated and, as the total sum of them, an integral representation of the Gibbs free energy of the system is derived. Difference of two free energies leads to an equation of state for osmotic pressure of small ion gas in an environment of macroions in the suspension.

\end{abstract}


\maketitle

\bigskip
In the Debye-H\"{u}ckel theory~\cite{Debye-Huckel} which is a prototype of mean field description, the Helmholtz and Gibbs free energies of an electrolyte, $F$ and $G$, are calculated in a self-consistent manner by linearizing the PB (Poisson-Boltzmann) equation. The difference $G\!-\!F$ being identical with the equation of state for osmotic pressure turns out to consist of a term of the ideal gas law, {\it i.e.}, the van't Hoff law, and a negative-definite term representing the effect of electric charges. Interpreting this decrease of the osmotic pressure as a manifestation of attractive nature of a net electric interaction and applying the DH (Debye-H\"{u}ckel) theory to macroionic systems, Langmuir~\cite{Langmuir} recognized first a possible existence of {\it a long-range Coulombic attraction between macromolecules with like electric charges} and predicted that the equation of state resembling a van der Waals isotherm leads to an instability and phase separation of ionic species. Verwey and Overbeek had criticized strongly this ingenious insight by pointing out {\lq\lq}the danger involved in the use of the linear approximation characteristic of the DH theory{\rq\rq}~\cite{VerweyOverbeek}.

Recent development of experimental techniques has revealed existence of inhomogeneities of particle distribution in colloidal suspensions such as clusters and voids~\cite{Dosho,CrockerGrier,ItoYoshidaIse,Yoshida}. These new large scale phenomena, along with well-known observations as salt-induced-melting of colloidal crystals~\cite{Hachisu}, crystalization through multi-stage phase transitions~\cite{SogamiYoshiyama} and reentrant solid-liquid transitions~\cite{Yamanaka}, stimulate strongly studies of charge-stabilized macroionic systems. Therefore, it is natural and reasonable to revisit the long-ignored idea of Langmuir which had predicted an occurence of phase separation in macroionic suspensions. The purpose of this article is to formulate a mean field theory for charge-stabilized suspensions of macromolecules~\cite{Sogami,SogamiIse} to revive the Langmuir's viewpoint without recourse to a linear approximation.

Direct application of the results of the DH theory to macroionic systems in the original Langmuir article~\cite{Langmuir} led necessarily to an unphysical {\it ionic democracy}~\cite{SogamiIse} where all the ionic species are treated so symmetrically that small ions must have surrounding clouds of macroions as well as macroions have clouds of small ions and small ions have mutual atmospheres of small ions with opposite charges. To remedy this apparent drawback of Langmuir's treatment, we adopt a viewpoint of {\it ionic aristocracy}~\cite{SogamiIse} where the size and the time scale of motion of macroions are considered to be asymmetrically larger than those of small ions. Postulating that macroions are adiabatically fixed, we derive integral representations of the Helmholtz and Gibbs free energies of macroionic suspension $F$ and $G$ in terms of an average electric potential satisfying the PB equation. The difference $G\!-\!F$ leads to the equation of state for the osmotic pressure of small ion gas in a fixed configuration of macroions.

We investigate a macroionic suspension with $M$ numbers of macromolecules in a container with a sufficiently large volume $\Omega$. Henceforth macromolecules are called particles for brevity. Effective values of the surface charge~\cite{Yamanaka}, volume and surface area of the $n$-th particle are $Z_n e$, $\omega_n$ and $S_n$, respectively. The gas of small ions with valency $z_j$ occupy the volume
\begin{equation}
     V = \Omega - \sum_{n = 1}^M \omega_n.
     \label{VOmegaomega}
\end{equation}
As a basic postulate of the present scheme, the thermodynamic function $Q$ of the system, {\it viz.}, the internal energy $U$, the Helmholtz free energy $F$ and the Gibbs free energy $G$, is generically postulated to consist of the sum of its electric part $Q^{\rm el}$ and non-electric part $Q^0$ as
\begin{equation}
   Q = Q^{\rm el} + Q^0.
\end{equation}

To make arguments clear, we model macroionic suspensions as follows:
\begin{enumerate}
\item Solvent of the system forms a uniform medium of background with a dielectric constant\,$\epsilon$\,.
\item Particles take a static distribution. The configuration of the particles is represented by the set $\{ \bR\,\} = \{ \bR_1 ,\bR_2, \cdots \bR_M \}$ of their center-of-mass coordinates.
\item In a static environment of particles，the gas of small ions reaches rapidly to a thermal equilibrium in which the gas of $j$-th species  of small ions with valency $z_j$ $(j=1,\cdots,s)$ has a number density given by the Boltzmann distribution
\begin{equation}
      n_{j}(\br) = n _ { j_0}e^{-z_j \phi(\br)}
      \label{Boltzmanndis}
\end{equation}
where $\phi(\br)$ is a ratio of the ordinary average elecrtic potential $\psi(\br)$ to an average thermal energy $kT = 1/ \beta$, {\it i.e.},
\begin{equation}
      \phi(\br) = \beta e \psi(\br).
\end{equation}
The coefficient $n_{j_0}$ in Eq.(\ref{Boltzmanndis}) is a constant determined by the number $N_j$ of small ions in the volume $V$ as
\begin{equation}
      N_j = n _ { j_0} \int_Ve^{-z_j \phi(\br)} dV.
      \label{Nsmallions}
\end{equation}
\item The dimensionless potential $\phi(\br)$ satisfy the PB equation
\begin{equation}
      \nabla^2 \phi(\br) = -4\pi \lambda_B\sum_{j=1}^s z_j n_j(\br)
      \label{PBeq}
\end{equation}
and the boundary conditions
\begin{equation}
      \bnu_n \cdot \nabla \phi(\br) = -4\pi \lambda_B Z_n \sigma_n(\br)
      \label{PBboundary}
\end{equation}
on particle surfaces\,$S_n(n = 1, \cdots , M)$\, where $\lambda_B$\,is the Bjerrum length $\lambda_B=e^2\beta/\epsilon$，$\bnu_n$ is a unit vector normal to the surface $S_n$ and $\sigma_n(\br)$ is a surface charge density normalized by the condition of surface integral
\begin{equation}
     \oint_{S_n} \sigma_n (\br) dS_n = 1.
\end{equation}
For brevity, Eq.(\ref{PBboundary}) is considered for $n=0$ to represent the boundary condition of the inner surface of the container $S_0$ which has a surface charge $Z_0e$, a surface charge density $\sigma_0({\br})$ and a unit normal vector $\bnu_0$.
\end{enumerate}

In analytical mechanics, the basic equations for various systems are described in a unified way by the Euler-Lagrange equation which is derived from the principle of minimum action. The macroionic suspensions are quite different from the mechanical systems. Nevertheless, it turns out possible to construct a generating fuctional from which the basic equation and the bounary conditions, {\it viz.}, Eqs. (\ref{PBeq}) and (\ref{PBboundary}), are deduced by imposing the extremal condition upon it. In fact，we can prove the following theorem.

{\bf [Theorem 1]}\ The generating functional for the PB equation and its boundary conditions is given by
\begin{eqnarray}
     \beta {\cal F}[\chi] & = & -{1 \over 8 \pi \lambda_B}\int_V 
        \{ \nabla \chi(\br)\} ^2 dV
       \nonumber \\
        & & + \sum_{n=1}^M Z_n \oint_{S_n} \chi(\br)\sigma_n(\br)dS_n
        \nonumber\\
        & &  - \sum_{j=1}^s N_j \ln \left({1 \over V} 
        \int_V e^{-z_j \chi(\br)} dV \right)
      \label{GeneratingF}
\end{eqnarray}
where $\chi({\br})$ is an unknown smooth function，the volume integral is taken over $V$ and the surface integrals are done on the surfaces of particles and the inner surface of the container.

To examine the extremal condition of the functional, let us calculate the difference between ${\cal F}[\chi+\delta\chi]$ and ${\cal F}[\chi]$. Applying the Gauss integral theorem, we obtain
\begin{eqnarray}
   \delta\beta{\cal F}&\!\!\equiv\!\!&\beta{\cal F}[\chi + \delta \chi]
         - \beta {\cal F}[\chi]
       \nonumber \\
        &\!\!=\!\!&\!{1 \over 4 \pi \lambda_B}\!
        \int_V\!\left(\!\nabla^2\chi + 4\pi\lambda_B\sum_j z_j
        \frac{N_j e^{-z_j\chi}}{\int_Ve^{-z_j\chi}dV}\!\right)
        \!\delta\chi dV
       \nonumber \\
        &\!\!+\!\!&  {1 \over 4 \pi \lambda_B} 
            \sum_n \oint_{S_n}(\bnu _n \cdot \nabla \chi
        + 4 \pi \lambda_B Z_n \sigma_n ) \delta \chi dS_n.
        \label{deltacalF}
\end{eqnarray}
For the functional ${\cal F}[\chi]$ to take an extremum at $\chi(\br)=\phi(\br)$，$\delta\beta{\cal F}$ must be an infinitesimal of second order with respect to an arbitrary variation $\delta\chi$ as
\begin{equation}
   \delta\beta{\cal F } 
   = \beta{\cal F }[\phi+\delta\chi]-\beta{\cal F }[\phi]
   = {\cal O}\delta(\chi^2).
   \label{condinfinitesimal}
\end{equation}
This implies that the right hand side of Eq.(\ref{deltacalF}) which consists of infinitesimals of the first order must vanish at $\chi(\br)=\phi(\br)$. Therefore, the extremal condition of the functional ${\cal F}[\chi]$ leads simultaneously to the PB equation (\ref{PBeq}) and the boundary conditions (\ref{PBboundary}) by using the relation between Eq.(\ref{Boltzmanndis}) and Eq.(\ref{Nsmallions}). For later arguments, it is convenient to express the extremal condition of the generating functional at $\chi = \phi$ in Eq.(\ref{condinfinitesimal}) as
\begin{equation}
\delta \beta{\cal F}[\phi]/\delta\phi=0.
  \label{extremalcondition}
\end{equation}
As shown below, the generating functional ${\cal F}$ is deeply related with the Helmholtz free energy of the system.

The electric part of the internal energy $U^{\rm el}$ is identical to the average value of electric energy $E$ of the system which is given, in the mean field description, by
\begin{eqnarray}
   \!\!\!\!\beta U^{\rm el} &\!\equiv\!& \beta E
            = {1 \over 8 \pi \lambda_B} \int_V(\nabla \phi )^2 dV
       \nonumber \\
           &\!=\!& {1\over 2} \sum_j \int_V z_j n_j \phi dV 
            + {1\over 2}\sum_n Z_n\!\oint_{S_n}\!\!\phi \sigma_n dS_n.
    \label{EelectricE}
\end{eqnarray}
The right hand side of this expression is obtained by using the Gauss integral theorem and the PB equation with associated boundary conditions. This estimation for $E$ is exact in the mean field description.

The Helmholtz free energy $F$ and the internal energy $U$ of the system are related by the themodynamic relation
\begin{equation}
\left(\frac{\pd \beta F}{\pd\beta}\right)_{V, N_j, M, \{\bR\,\}} = U
   \label{thermodyidentity}
\end{equation}
where the derivative with respect to $\beta$ is taken for the fixed values of the volume，the numbers of small ions and particles and also for the fixed configuration of particles. In the present scheme, the same relations are assumed to hold separately both for the electric parts $F^{\rm el}$ and $U^{\rm el}$ and the non-electric parts $F^0$ and $U^0$ of the Helmholtz free energy and the internal energy.

Here we use the thermodynamic relation in Eq.(\ref{thermodyidentity}) for $F^{\rm el}$ and $U^{\rm el}=E$ to find the following theorem.\par
{\bf [Theorem 2]} The electric part of the Helmholtz free energy of the system $F^{\rm el}$ is the extremum of the generating functional, {\it i.e.}, $F^{\rm el} = {\cal F}[\phi]$.

To prove this theorem we notice that ${\cal F}[\phi]$ depends on $\beta$ through the Bjerrum length $\lambda_B$ and the average electric potential $\phi$. It is readily shown that
\begin{eqnarray}
   \noalign{\vskip 0.1cm}
   \left(\frac{\pd \beta {\cal F}[\phi]}{\pd\beta}\right)
   &=& \frac{\pd \lambda_B}{\pd\beta}
    \left(\frac{\pd \beta {\cal F}[\phi]}{\pd \lambda_B}\right)_\phi
   + \frac{\pd \phi}{\pd\beta}
    \left(\frac{\delta \beta {\cal F}[\phi]}{\delta \phi}\right)_{\lambda_B}
   \nonumber \\
   \noalign{\vskip 0.1cm}
   &=& kT\frac{1}{8\pi\lambda_B}\int_V\left\{(\nabla\phi(\br))^2\right\}dV
   \nonumber \\
   &=& E
   \label{betaderivative}
\end{eqnarray}
where the condition of extremum in Eq.(\ref{extremalcondition}) is used. As is readily confirmed it is possible to prove this theorem also by using the Debye charging-up formula.

In this way the Helmholtz free energy of the macroionic suspension $F$ is expressed in terms of the average electric potential $\phi$ satisfying the PB equation as follows:
  \begin{eqnarray}
          F &=& {\cal F}[\phi] + F^0
            \nonumber \\
            \noalign{\vskip 0.1cm}
            &=& - {1\over 2}kT\sum_j z_j \int_V n_j \phi dV
             + {1\over 2}kT\sum_n Z_n \oint_{S_n} \phi\sigma _n dS_n
            \nonumber \\
            & & - kT\sum_j N_j \ln \left ({1 \over V}
             \int_V e^{-z_j \phi} dV \right) + F^0.
   \label{Helmholtz}
  \end{eqnarray}
Notice that this integral representaion is exact in the mean field description. If we use the identical relation in Eq.(\ref{EelectricE}), the Helmholtz free energy $F$ can take various different but equivalent forms being convenient for each purpose.

Integration of the PB equation in Eq.(\ref{PBeq}) over the volume $V$ and usage of the Gauss integral theorem for the boundary condition in Eq.(\ref{PBboundary}) result in the condition of electric neutrality
\begin{equation}
      \sum_j z_j N_j + \sum_n Z_n = 0.
      \label{neutralcon}
\end{equation}
In the macroionic suspension where the number and configuration of particles are fixed, this condition requires that the effective valency of particles $Z_n$ must change against a variation of the number of small ions $N_i$. Therefore the chemical potential of the small ions which is obtained by $\pd F/\pd N_j$ must necessarily be correlated to the derivative of $F$ with respect to $Z_n$. It is reasonable to interpret the derivative of the Helmholtz free energy $\pd F/\pd Z_n$ as the {\it chemical potential of effective valency of the $n$-th particle}~\cite{Sogami,SogamiIse}. The Gibbs free energy of the system $G$ is obatined by summing up the chemical potentials of the small ions of all kinds and the chemical potentials of the effective valencies of all particles.

To take derivatives of $F$ with respect to $N_j$ and $Z_n$ and to sum up the component chemical potentials without contradiction to the constraint condition~\cite{Dirac}, we introduce the following differential operator as
\begin{equation}
     D_{NZ} = \sum_j N_j\frac{\pd}{\pd N_j} + \sum_n Z_n\frac{\pd}{\pd Z_n}.
\end{equation}
It is readily confirmed that, when applied to the neutrality condition, this operator reproduces the condition again without resulting in any additional constraint~\cite{Dirac}. Therefore, this operator $D_{NZ}$ can be interpreted as a kind of {\it covariant derivative} which preserves invariance of a surface spanned by the constraint in a fictitious space $(N_1,\cdots,N_s, Z_1,\cdots,Z_M)$.

For application of the covariant derivative $D_{NZ}$, it is convenient to use the Helmholtz free energy in the form
\begin{eqnarray}
     \beta F & = & -{1 \over 8 \pi \lambda_B}\int_V
        \{ \nabla \phi(\br)\} ^2 dV +
        \sum_n Z_n \oint_{S_n} \phi(\br)\sigma_n(\br)dS_n
        \nonumber\\
        & & - \sum_j N_j \ln \left({1 \over V} 
        \int_V e^{-z_j \phi (\br)} dV \right) 
        + \beta F_0.
      \label{HelmholtzF}
\end{eqnarray}
Noting that the average potential $\phi$ as the solution of the PB equation depends on $N_j$ and $Z_n$, we find
\begin{eqnarray}
  D_{NZ}(\beta F)
  &=& \sum_j N_j\left(\frac{\pd \beta F}{\pd N_j}\right)_{Z_n, \phi}
   + \sum_n Z_n\left(\frac{\pd \beta F}{\pd Z_n}\right)_{N_j, \phi}
   \nonumber\\
   \noalign{\vskip 0.1cm}
   & & + D_{NZ}(\phi)
         \left({\delta\beta F \over\delta\phi}\right)_{N_j, Z_n}.
\end{eqnarray}
Then, using the extremal condition in Eq.(\ref{extremalcondition}) and the definition $G=D_{NZ}(F)$, we obtain the theorem.\par
{\bf [Theorem 3]} The Gibbs free energy of the macroionic suspension has the integral representation
\begin{eqnarray}
\!\!\!
    G &=& kT\sum_n Z_n \oint_{S_n} \phi(\br)\sigma_n(\br)dS_n
      \nonumber \\
      & & - kT\sum_j N_j
          \ln \left({1 \over V}\int_V e^{-z_j \phi(\br)} dV\right)
        + G_0 
      \label{GibbsG}
\end{eqnarray}
where $G_0= D_{NZ}(F_0)$ is a non-electric part of the Gibbs free energy.

Among the three integral representaions in Eqs.(\ref{EelectricE}), (\ref{HelmholtzF}) and (\ref{GibbsG}), there holds the relation
\begin{equation}
   G - G_0 = F - F_0 + E.
\end{equation}
On the other hand, the difference between $G$ and $F$ is related to the osmotic pressure of the suspension as
\begin{equation}
   G - F = (P_0 + P)V
\vspace*{-0.2cm}
\end{equation}
which reduces to
\begin{equation}
   G_0 - F_0 = P_0V + kT\sum_j N_j
\vspace*{-0.2cm}
\end{equation}
at the limit where all the solutes become neutral. Here，$P_0 V$ and $kT\sum_j N_j$ are contributions from the pressure term of solvent and the van't Hoff term. Therefore, the gass of small ions in the environment of macroions is subjected to the equation of state
\begin{equation}
    PV = kT\sum_j N_j + E
\vspace*{-0.2cm}
    \label{equationofstate}
\end{equation}
which is exact in the mean field description.

In this way we have derived the equation of state for osmotic pressure of small ions without resort to the linear approximation. The right hand side of Eq.(\ref{equationofstate}) includes the average electric energy $E(\{\bR\,\})$. In contrast to the equation of state in the DH theory of a simple electrolyte where the net contribution of electric charges is a negative-definite constant, the average electric energy $E(\{\bR\,\})$ takes various values depending on the configurations of particles. Which sign does $E$ take? It is safely argued that $E$ is negative for almost all thermodynamically equilibrium states. To testify to this assertion, it is sufficient to apply the condition of electric neutrality to an arbitrary region with medium size in the suspension. Ions with unlike charges tend to come closer and ions with like charges tend to push farther apart in average so that the condition of electric neutrality is maintained in the region. As a result the net electric energy becomes negative. The electric energy may be instantaneously positive in a region in a suspension where particles take an extremely singular configuration. Such a configuration, however, is not thermodynamically stable.

Therefore it is possible to conclude that, although the mathematical formulation of his theory was immature, the insight of Langmuir was correct. The simple form of Eq.(\ref{equationofstate}) is a direct manifestation of the Langmuir's hypothesis of the Coulombic attraction. An electrically stable configuration of particles with like charges which is established through the intermediary of small ions among particles tends inevitably to attract and cage the small ions, and consequently reduces the osmotic pressure. The simple form of Eq.(\ref{equationofstate}) enables us to proceed computer simulations. Notice that, if the equation of state includes, not the electric energy $E$, but a free energy $F$ or $G$, its physical implication becomes much more indirect and computer simulations becomes difficult.

In the mean field theory with the linearized PB equation~\cite{Sogami,SogamiIse}, the equation of state in Eq.(\ref{equationofstate}) had been derived and the approximate Gibbs free energies had been proved to have a medium-range strong repulsive part and a long-range attraction tail. Thermodynamic processes of particles develop slowly toward a direction in which the value of the Gibbs free energy $G(\{\bR\,\})$ decreases. It is the long-range attractive tail of the pair effective interaction included in $G(\{\bR\,\})$ that is interpreted to promote the formation of large scale phenomena such as clusters and voids. In this connection it is worthwhile to point out that the Langmuir theory is related with volume-term theories~\cite{Hansen,Warren} which is now advocated to explain phase separations in macroionic suspensions by a spinodal instability sensitive only by small ions. Schmitz et al.~\cite{SchmitzBhuiyan,Schmitz} clarified mutual relationships among the Langmuir theory, the theory with the Gibbs attractive interaction and the volume-term theory essentially within the extent of linearized mean field description. The present formalism of mean field theory without linear approximation strengthens validity of their arguments.

This study is partially funded by {\lq\lq}Ground Research for Space Utilization{\rq\rq} promoted by NASDA and Japan Space Forum.

\end{document}